\documentclass{article}
\usepackage[margin=1in]{geometry}
\usepackage{graphicx} 
\usepackage{amssymb}
\usepackage{amsmath,esint}
\usepackage{subcaption}
\usepackage{authblk}

\title{Cross-Relation Characterization of Knowledge Networks}

\author[+,*]{ Eric K. Tokuda}
\author[*]{ Renaud Lambiotte}
\author[+]{ Luciano da F. Costa}

\affil[*]{Mathematical Institute, University of Oxford, OX2 6GG Oxford, United Kingdom}
\affil[+]{Institute of Physics at S\~ao Carlos, University of S\~ao Paulo, \authorcr
 Avenida Trabalhador São-carlense, 400, 13566-590
São Carlos, S.P., 1356, Brazil}
\date{February 2023}

\begin{document}
\maketitle

\begin{abstract}
Knowledge networks have become increasingly important as a changing repository of data which can be represented, studied and modeled by using complex networks concepts and methodologies.  Here we report a study of knowledge networks corresponding to the areas of Physics and Theology, obtained from the Wikipedia and taken at two different dates separated by 4 years. The respective two versions of these networks were characterized in terms of their respective cross-relation signatures, being summarized in terms of modification indices obtained for each of the nodes that are preserved among the two versions.  The proposed methodology is first evaluated on Erd\H{o}s–Rényi (ER) and Barabási–Albert model (BA) networks, before being tested on the knowledge networks obtained from the Wikipedia respectively to the areas of Physics and Theology. In the former  study, it has been observed that the nodes at the core and periphery of both types of theoretical models yielded similar modification indices within these two groups of nodes, but with distinct values when taken across these two groups.  The study of the Physics and Theology networks indicated that these two networks have signatures respectively similar to those of the BA and ER models, as well as that higher modification values being obtained for the periphery nodes, as compared to the respective core nodes.
\end{abstract}

\section{Introduction}

One important characteristic of real-world structures is that the properties of the respective components are shared or relate to one another.  For instance, species of plants within a same family will intrinsically share several structural and functional features.  A similar situation can be observed for words in language, which have their meanings interrelated in several manners.  Yet another important example consists of knowledge, where each concept is inherently connected with other concepts along a respective ontological framework.

As a consequence of the inherent ability of graphs and complex networks for representing relationships between a given set of entities \cite{newman2018networks}, these structures have been applied to representing, visualizing, characterizing and modeling a wide range of real-world as well as abstract complex systems.   In the particular case of knowledge, for instance, each of its portions can be compartmentalized and represented as a respective node, while the interrelationships between these concepts can be effectively expressed in terms of respective links, or edges.  This abstraction leads to a complex network that can then be studied from several perspectives while taking into account the respective interrelationships.  

Given a body of knowledge, such as a scientific area represented in some respective framework (e.g.~books, encyclopedia, Wikipedia), this area can be subdivided into several respective portions associated to major concepts which are then represented as respective nodes.  For instance, in the case of Physics, possible sub-areas could include ``Thermodynamic limit'', ``Supersymmetry'' and ``String theory''.  These nodes can then be interconnected while taking into account the respectively shared concepts/words or by considering the respective cross-references in the original document.  Though the present work focuses on non-directed networks, the reported concepts and methods can be readily adapted to directed knowledge networks.

Cross-reference based knowledge networks, in which nodes
correspond to portions of a document and references between
these portions are represented as respective edges, have been often studied in the literature (e.g.~\cite{silva2011investigating,farber2018linked,chen2020review}).  
While this type of networks can provide comprehensive information about the interrelationships between the involved concepts, the more systematic quantitative characterization of the respective topology has often been approached by using measurements including node degree, clustering coefficient, shortest paths, etc. (e.g.~\cite{costa2007characterization}), as well as respective concentric versions (e.g.~\cite{da2008concentric,silva2011investigating,travenccolo2008hierarchical}).

Described more recently~\cite{da2022autorrelation}, the concept of \emph{cross-relation} aims at characterizing the topological properties of each specific node $i$ in a given network in terms of a respective \emph{signature} extending along successive \emph{lags} or \emph{shifts} through the neighborhoods of node $i$, in a manner that is analogous but not equivalent to the more traditional concept of cross-correlation.  More specifically, for each given node $i$, its hierarchical neighborhoods at successive topological distances $\delta$ are identified.  Then, the similarities between the topological properties of the reference node and those of each of the nodes in the successive neighborhoods are estimated, and the respective average is taken as an indication of the overall topological similarity between the reference node $i$ and the nodes at the neighborhood at distance $\delta$.  Whole signatures can then be obtained respectively to each specific node in the given network.  

The estimation of cross-relation between two networks requires at least a portion of the nodes to be \emph{aligned} (or matched).  By aligned it is henceforth understood that a pair of nodes refer, from each of the two networks, refers to the same entity, though their interconnections in the respective networks can vary.  In this case, the respectively obtained signature, understood to correspond to the \emph{cross-relation} between each pair of aligned nodes in the networks, can be obtained and employed as a subsidy for comparing the topological properties of each node along its successive hierarchical levels.  
As an example, given two knowledge networks obtained at different dates, the nodes referring to the same concepts can be readily identified as aligned nodes, which allow them to be compared in terms of the cross-relation signatures.  
The effectivity of the two aforementioned concepts, in the sense of achieving more strict and detailed comparison, stems from the properties of the recently introduced coincidence similarity index and respectively derived networks~\cite{costa2021coincidence,costa2023multiset}, which provide a particularly selective resource for comparing networks, especially when they are similar.

A particularly interesting case consists of networks that change along time. In this case, a pair of aligned nodes (knowledge entities) would correspond to the same node in two instants of time. Comparing the entire network using the proposed approach may be impractical, depending on the size of the network. A commonly adopted approach is to focus on subnetworks, such as the common divisions of science (e.g. Humanities, Life Sciences), or narrower fields (e.g. Mathematics, History).

In general, network comparison can be performed by focusing on their similarity at different scales \cite{donnat2018tracking}. At the meso-scale, networks can be compared respectively to their community structure~\cite{fortunato2010community}. This comparison can be done in several ways, e.g.~one can compare the count of links within and between communities~\cite{porter2009communities} or use the statistical properties of Markov processes~\cite{lambiotte2014random}. These patterns characterizing the internal communities can then be used to group networks. In~\cite{onnela2004clustering} the authors described a taxonomy of a set of real-world networks and noticed that the groupings agreed with their respective categories (e.g. financial, road). Methods focusing on the global scale focus instead on spectral information~\cite{mellor2019graph,edmunds2018spectral}. In~\cite{mellor2019graph}, network structures are compared using the distribution of eigenvalues of non-backtracking random walks. The method allows consideration of networks of different sizes (number of nodes and edges), being relatively fast in terms of computation time. The application to synthetic and real-world networks showed that it could efficiently cluster networks of the same types.

The current work aims at characterizing knowledge networks obtained from the Wikipedia  using the cross-relation methodology.  Given a pair networks to be compared, topological node features are extracted for each aligned node, the cross-relation method is applied and signatures from the nodes are obtained.  Then, the absolute difference between the signatures across aligned nodes are calculated, and the sum of the differences across lags are used as our final metric, the modification index, which accounts for the modification respectively undergone by the two networks along successive neighborhoods of each aligned node.   The network can then be visualized by scaling the node sizes proportionally to the modification index.

The proposed method has been applied to two distinct problems: to analyze theoretical-model networks, and to analyze Wikipedia networks considered at two time instants.  Three theoretical models have been considered (ER, BA and WS). The results confirmed the effectiveness of the cross-relation approach in identifying the topological changes around aligned nodes undergone by the reference and modified versions of the same network.  

The cross-relation method was then applied to identify the changes of the topology around aligned nodes considering knowledge networks (Physics and Theology) derived from the Wikipedia. In addition, the cross-relation approach succeeded in identifying the nodes that underwent the more intense modifications in the networks considered between the two considered time instants.  
The consideration of successive neighborhoods in the cross-relation signatures allowed enhanced characterization of the topological changes, including cases in which the changes took place at nodes further away from the reference nodes.  Of particular interest is the indication that the topology around those nodes closer to the border of the networks tended to be more strongly influenced by changes along time, which provides subsidies for making link and node predictions.

The manuscript is organized as follows: we start by presenting the Wikipedia data and how the networks were obtained from it.  Next, the main theoretical concepts used in this work are described; which is then followed by the experiments and results concerning theoretical models and Wikipedia data.

\section{Dataset}
\label{sec:dataset}
The network obtained from Wikipedia represents the pages as nodes and the hyperlinks among them. Every page may have links (taken as being undirected) to multiple pages. The original network data was obtained from \cite{consonni2019wikilinkgraphs}. We considered two subnetworks of this network related to two scientific disciplines: \emph{Physics} and \emph{Theology}. In Wikipedia, some pages corresponding to complex topics have a corresponding \emph{category page}\footnote{https://en.wikipedia.org/wiki/Category:field}, which contains a list of subconcepts of that topic. We consider the category pages of Physics and Theology in order to find the most relevant pages in each field. Here, we considered the data in two moments 01 March 2014 and 01 March 2018.  

\section{Basic Concepts}
Here we describe the main respective to methodology applied in this work.

Principal component analysis (PCA) is a non-parametric statistical method commonly used for dimensionality reduction~\cite{gewers2021principal,costa2020acompact}. This method works by first determining the principal components, which are variables corresponding to linear combinations of the original data and that explain its maximal variance. These components form an orthonormal basis, onto which the original data can be projected to. In data analysis, and in particular in dimensionality reduction, only the first few principal components corresponding to the largest variance are typically considered.

Recently, several real-world problems have been approached using complex networks~\cite{costa2011analyzing}. This abstraction focuses on entities of interest, represented by nodes, and links implementing connections among them~\cite{costa2007characterization}. A node connected to \emph{k} other nodes is said to have degree $k$. A network of $n$ nodes and a set of corresponding $\{k_i\}$ degrees, has an average degree that can be calculated as:

\begin{align}
\langle k \rangle = \frac{1}{n} \sum_{i=1}^n k_i
\end{align}

Several topological models of networks have been described in the literature (e.g.~\cite{costa2007characterization}). The Erd\H{o}s–Rényi model (ER)~\cite{erdHos1960evolution} constitutes a simple random structure specified by the number of nodes $n$ and the adopted interconnecting probability. In this model, all links are equiprobable and as such all possibilities of networks with the same set of vertices and number of links are equally likely.~\cite{costa2007characterization}.
The Watts–Strogatz model (WS) is a more complex random network model which adds a \emph{rewiring} procedure into the network construction. This network model has a rewiring parameter $\beta$. It starts with a ring lattice of desired size and each of the links connecting a given node to other nodes has its destination changed with a given probability $\beta$, while preventing self-loops and double links.  The clustering coefficient and shortest path lengths of the obtained networks depend on the value of $\beta$.
The Barabási–Albert model (BA) employs preferential attachment during the network creation. More specifically, nodes are progressively added with connections favouring higher degree nodes. One of the main differences of BA networks and the previous two models concerns the vertex degree distribution, which exhibits a long tail, leading to a power law degree distribution.

Visualizing networks represents a demanding problem, which requires a trade-off among discernibility, depiction and performance~\cite{silva2018visualizing}. Force-directed network visualizing algorithms, such as  \emph{Fruchterman-Reingold}~\cite{fruchterman1991graph}, are based on a system of physical forces that iterates until the mechanical equilibrium is achieved.

When comparing traditional sets or vectors (both taken as being non-zero), one important matter concerns how to define that two sample are ``similar'' to one another. Various similarity indices have been proposed for that finality (e.g.~\cite{wolda1981similarity}), including the widely used Jaccard and the interiority indices (e.g.~\cite{costa2021further}). While the Jaccard has been widely adopted in scientific domains, it does not indicate to which extent one set falls inside the other~\cite{costa2021onsimilarity}. Suggested recently~\cite{costa2021coincidence}, the coincidence similarity addresses this problem by combining the Jaccard and the interiority indices.

Similarity indices are not constrained to sets, often being applicable also to real-valued vectors or functions~\cite{costa2021multiset}. More specifically, the coincidence similarity of two non-zero functions $\boldsymbol{f}$ and $\boldsymbol{g}$ is given by:

\begin{equation}
  \mathcal{C}_R(\boldsymbol{f},\boldsymbol{g}) =  \mathcal{I}_R(\boldsymbol{f},\boldsymbol{g}) \  \mathcal{J}_R(\boldsymbol{f},\boldsymbol{g})
\end{equation}

where $\mathcal{I}$ and $\mathcal{J}$ are the interiority and the Jaccard coefficients, given as:

\begin{equation}
  \mathcal{I}_R(\boldsymbol{f},\boldsymbol{g}) = \frac{ \sum_i \min\left\{ | f_i |, | g_i |\right\}}
  {\min\left\{ \sum_i | f_i |, \sum_i | g_i | \right\}}
\end{equation}

\begin{equation}
  \mathcal{J}_R(\boldsymbol{f},\boldsymbol{g}) = \frac{ \sum_i \min\left\{ | f_i |, | g_i |\right\}}
  {\max\left\{ \sum_i | f_i |, \sum_i | g_i | \right\}}
\end{equation}

An extended version of the coincidence index allows the incorporation of a parameter $D$, $D\in \mathbb{N}$, for a control of the strictness of the comparison. An analysis of the influence of $D$ is presented in~\cite{costa2021multiset}.

\begin{equation}
    \mathcal{C}_R({\boldsymbol f},{\boldsymbol g}) = 
     \mathcal{I}_R({\boldsymbol f},{\boldsymbol g}) \ \mathcal{J}_R({\boldsymbol f},\boldsymbol{g}) ^D
\end{equation}

The concept of similarity can also be applied to the study of complex networks. Given two nodes of a network and corresponding measurements, they can be compared using the previously explained coincidence index. This allows \emph{coincidence similarity networks} to be obtained~\cite{costa2021coincidence}. The edges of these obtained networks  represent the coincidence similarity between the features of the respective pair of nodes. Thanks to the enhanced selectivity and sensitivity of the coincidence similarity, more detailed networks can be obtained by this methodology, which has been successfully applied to several situations (e.g.~\cite{costa2021onsimilarity,domingues2022identification,costa2023multiset}).

A fundamental interest in descriptive statistics concerns the quantification of how much two random variables tend to vary together, which is frequently estimated in terms of the measurements of covariance as well as respectively normalized \emph{cross-correlations} ~\cite{degroot2012probability,ross2014introduction}. The particular case in which the correlation refers to pairs of lagged observations along a same random variable is called \emph{auto-correlation}~\cite{degroot2012probability,ross2014introduction}. 

Analogously, considering the interesting advantages of the coincidence similarity index, the concepts of \emph{autorrelation} and \emph{cross-relation} have been introduced more recently~\cite{da2022autorrelation} in order to obtain more strict and complete characterizations of the relationships between the topological features of a reference node and the nodes at successive neighborhoods. The autorrelation signature (a vector) of a node incorporates the coincidence indices between the topological features of that node and all its lagged neighbours.  When applied to two networks, the cross-relation of two respective networks quantifies the coincidence between the adopted topological properties of a node in one of the networks and its counterpart neighbors in the other network~\cite{da2021comparing,da2022autorrelation}.

\begin{figure}
    \centering
    \includegraphics[width=.95\textwidth]{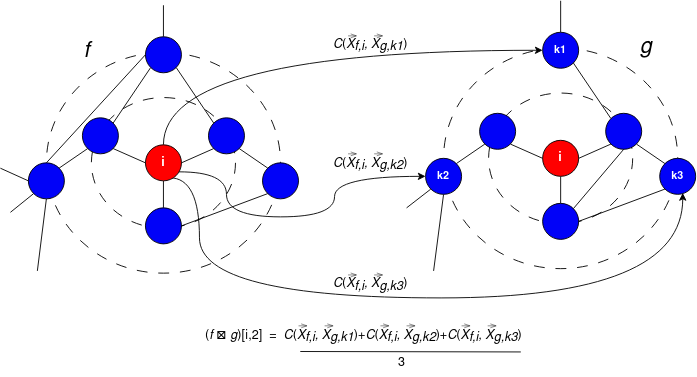}
    \caption{Cross-relation between two hypothetical networks $f$ and $g$, taking node $i$ as reference. More specifically, $\left( f \boxtimes g \right)\left[ i,2 \right]$ is calculated for the reference node $i$ with lag $\delta=2$ by averaging the coincidence values between the features of the reference node in network $f$, $X_{f,i}$, and the features of the respective neighbours in the topological layer 2 of network $g$, i.e.~$X_{g,k}$.}
    \label{fig:diagcrossrel}
\end{figure}

Consider two networks $f$ and $g$ with nodes in respective correspondence (node alignment), as illustrated in Figure~\ref{fig:diagcrossrel}.
Each node is characterized by associated features $x_j$, $j=1,2,\ldots,S$,  represented as
a respective feature vectors, for each network $f$ and $g$, as follows:

\begin{equation}
\vec{X}_{f,i} = \left[ x_{f,1,i} \quad x_{f,2,i} \quad \ldots x_{f,S,i} \right]^T
\end{equation}

\begin{equation}
\vec{X}_{g,i} = \left[ x_{g,1,i} \quad x_{g,2,i} \quad  \ldots x_{g,S,i} \right]^T
\end{equation}
\noindent where $x_{f,i,j}$ corresponds to the value of the feature $x_j$ of node $i$ in the network $f$. The cross-relation respective to lags $\delta=1,2,\ldots,\Delta$ and $|\mathcal{L}_{\delta}(i)|$ meaning the number of nodes in the layer $\delta$ relative to the node $i$, can then be calculated as:

\begin{equation}
\left( f \boxtimes g \right)\left[ i,\delta \right] = \frac{1}{|\mathcal{L}_{\delta}(i)|} \sum_{k \in \mathcal{L}_{\delta}(i)} C(\vec{X}_{f,i},\vec{X}_{g,k})
\end{equation}
For the specific case where the network is the same, $f=g$, the auto-relation is calculated as:

\begin{equation}
\left( f \boxtimes f \right)\left[ i,\delta \right] = \frac{1}{|\mathcal{L}_{\delta}(i)|} \sum_{k \in \mathcal{L}_{\delta}(i)} C(\vec{X}_{f,i},\vec{X}_{f,k})
\end{equation}

A direct application of such measurements concerns the comparison of networks. By averaging the cross-relation signatures across corresponding network nodes, an average \emph{network} signature is obtained, which can be used to compare networks. This idea has been proposed and explored in~\cite{da2022autorrelation}, where three theoretical models -- ER, BA, WS -- were compared by generating a set of networks with similar parameters for each model, extracting the network signature, calculating the derivative of the network signatures and then performing a principal component analysis~\cite{gewers2021principal} on the generated results to visually assess the separation. As a result, three quite distinct groups were observed corresponding to the three considered network models, which corroborated the potential of this approach to discriminate networks.

\section{Methodology}

The potential of the cross-relation to compare two networks has been preliminarly explored in~\cite{da2022autorrelation}. Here, we apply this effective approach to analyze the Wikipedia network at two instants along time.

\begin{figure}[ht]
  \centering
 \includegraphics[width=0.85\textwidth]{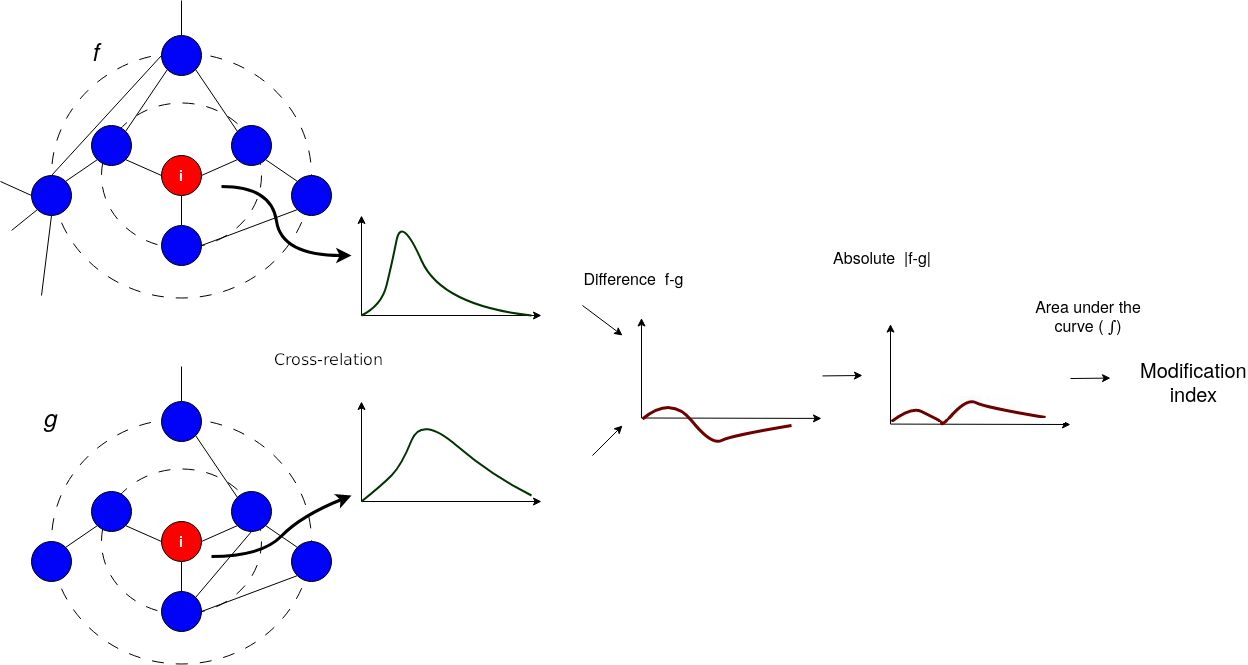}
   \caption{Illustration of the calculation of the modification index respectively to a node
   $i$ and two instances $f$ and $g$ of the networks to which it belongs to.}
  \label{fig:diagram2}
\end{figure}

The adopted method considers a \emph{modification index} for each node of the two networks analyzed. From each network (A and B), we initially extract node features (e.g.~degree, clustering coefficient). For each aligned node, i.e., that is both present in the two networks, we calculate the node cross-relation between the two networks (i.e.~\emph{A cross-related with B} and \emph{B cross-related with A}) for a fixed range of lags. As a result, two sets of same size of cross-relation signatures with the same length are obtained. For each aligned node, the absolute difference between the signatures is calculated and the sum of these differences is defined as the final modification index.   The calculation of the modification index is illustrated in the diagram in Figure~\ref{fig:diagram2}.

While the modification index is calculated only for the aligned nodes, the entire network is considered for the calculation of the adopted measurements.  Observe that, as a consequence of the modifications undergone by a network along time, including the addition and exclusion of nodes, not all nodes will remain aligned.  

An alternative modification index can be obtained analogously to the above scheme, but by taking into account the autorrelation signatures of the respective nodes in each of the two networks, instead of the respective cross-relation signatures.  Another possible approach is to weight the cross-relation signatures with weight values that decrease or increase along the hierarchical neighborhoods, depending on specific research questions.

\begin{figure}[ht]
  \centering
 \includegraphics[width=0.48\textwidth]{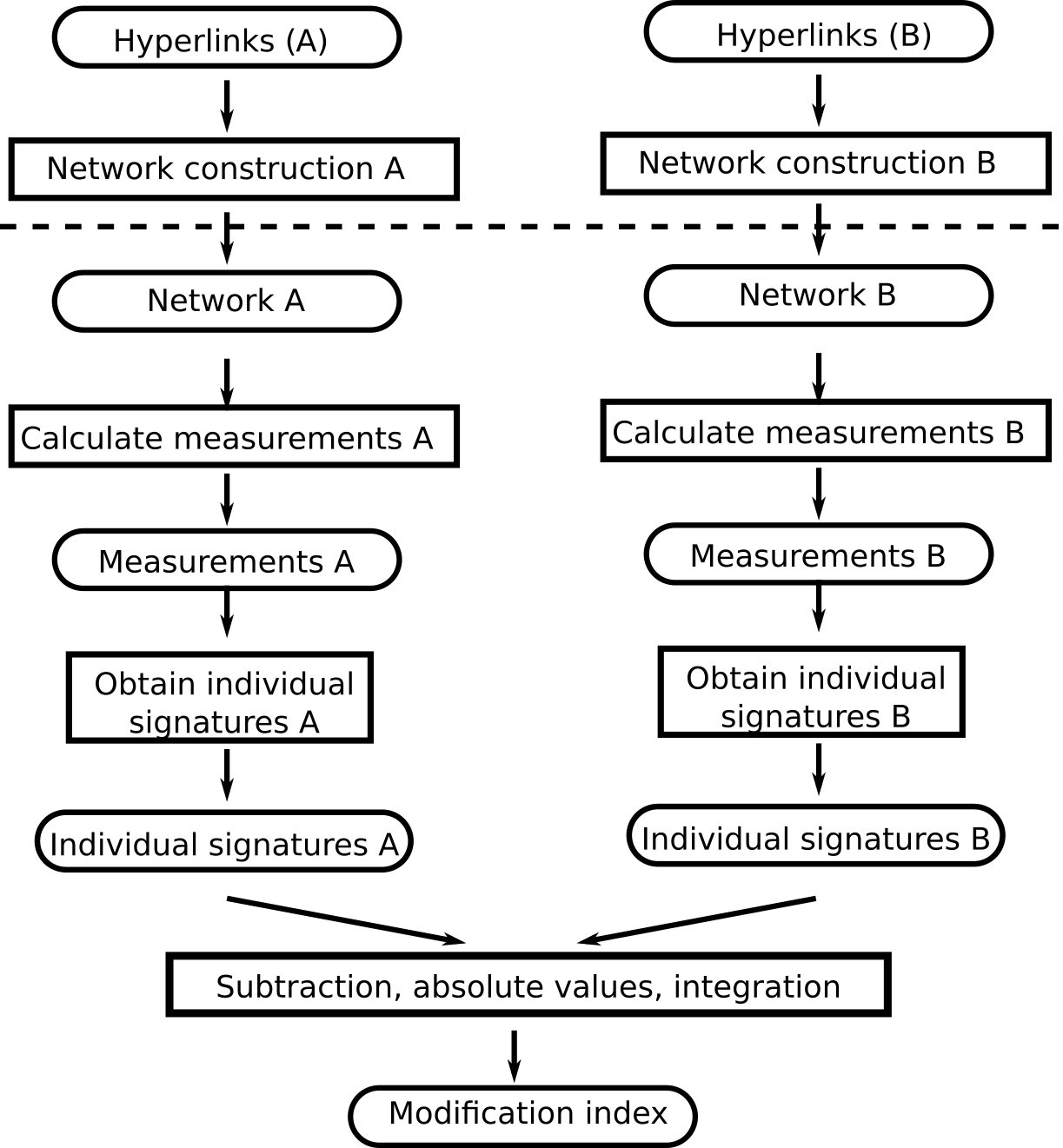}
   \caption{Diagram describing the main data and stages involved in the calculation of the the modification index of each node, given two instances A and B of a network.  In the cases where the networks A and B are already available, the processing start below the dashed lines, involving the calculation of the adopted topological measurements, followed by the determination of the individual cross-relation signatures between both networks.  The latter signatures are then subtracted, and the sum of the absolute values yields the modification indices for each node.    In case the initial dataset consists of hyperlinks (or references) between modules of text, as is the case with the knowledge networks considered in Section~\ref{sec:dataset}, the processing starts above the dashed line, including the construction of the networks from the respectively supplied hyperlinks.}
  \label{fig:diagram1}
\end{figure}

The visualization of the modification index obtained for each aligned node in the respective networks can be implemented in two main forms.  In case the changes are relatively large, the modification index values can be mapped linearly into the diameter of the respective nodes, e.g.~by using:
\begin{align}
    & x_{norm} = \frac{\mu - \min(\mu)}{\max(\mu) - \min(\mu)} \nonumber\\
    & sz = (x_{norm} * szmax) + szmin  
    \label{eq:linear}
\end{align}

\noindent where we henceforth adopt $szmin=5$ and $szmax=15$.
However, in  cases involving more subtle variations of the modification index, it is interesting to consider a logarithmic mapping such as:
\begin{align}
    & x = \log(\mu) \nonumber\\
    & x_{norm} = \frac{x - \min(x)}{\max(x) - \min(x)} \label{eq:log}\\
    & sz = (x_{norm} * szmax) + szmin \nonumber 
\end{align}

\section{Cross-Relation Between Theoretical Models}
\label{sec:theo}

Henceforth, we will adopt the Fruchterman-Reingold method~\cite{fruchterman} for visualizing the networks.  This is a force-directed methodology for assigning 2D positions to network nodes that tries to distribute the nodes as widely as possible (nodes are assumed to repel one another) while taking into account the respective interconnections (understood as attraction fields).  As a consequence of these enforced characteristics, the nodes with smallest degree tend to result along the periphery of the visualized networks.

\begin{figure}[ht]
     \centering
     \begin{subfigure}[b]{0.4\textwidth}
         \centering
 \includegraphics[width=\textwidth]{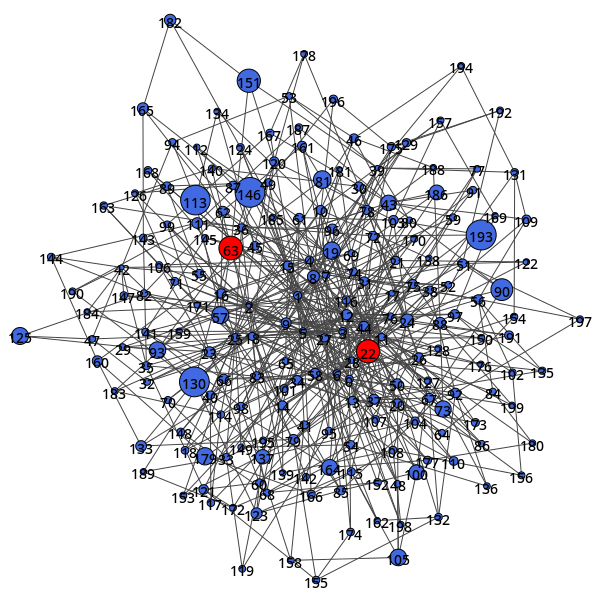}  
         \caption{}
     \end{subfigure} \qquad
     \begin{subfigure}[b]{0.4\textwidth}
         \centering
 \includegraphics[width=\textwidth]{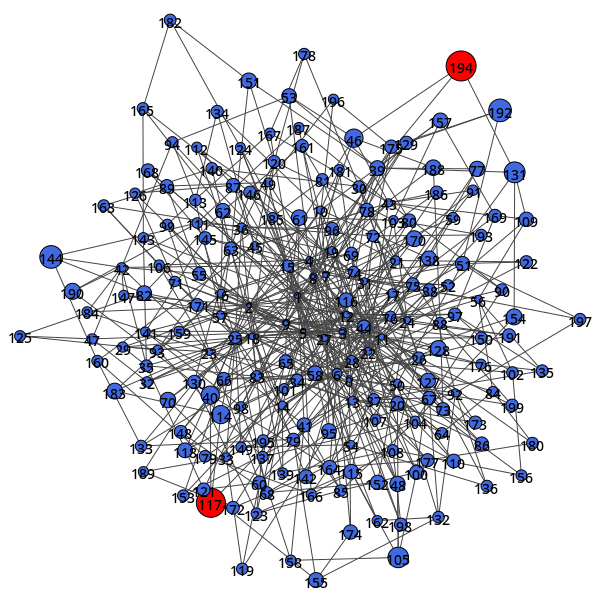}
         \caption{}
     \end{subfigure} \\
     \begin{subfigure}[b]{0.4\textwidth}
         \centering
 \includegraphics[width=\textwidth]{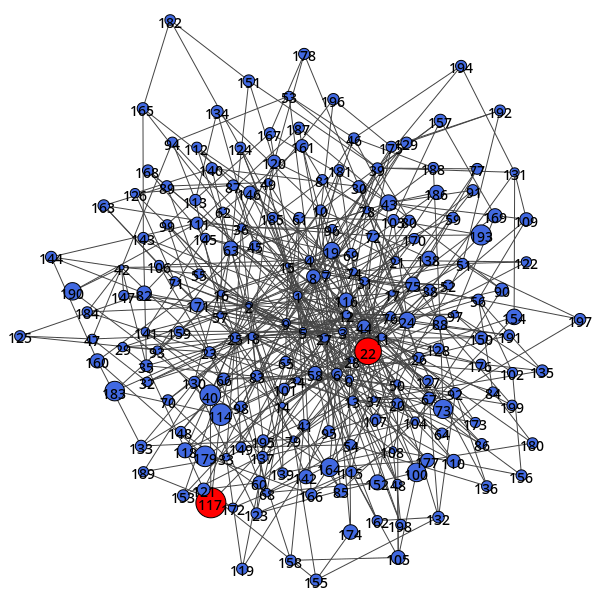}  
         \caption{}
     \end{subfigure}
   \caption{Modification index with the addition of a link. Three different scenarios were considered, which takes into account the positioning of the node in the force-directed disposition of node, whether it is a core or a border node. In (a), a link is added between two core nodes; in (b), between a core and a border node; and in (c), between two border modes.}
  \label{fig:handpicked}
\end{figure}

In Figure~\ref{fig:handpicked}, we can visualize the impact of modifying the network on the proposed measurement. More specifically, we analyzed how the modification indices changed with the addition of a single link. Three different scenarios were considered, taking into account whether the nodes that received the new link belonged to the core or border of the respective network. Hence, we considered three different scenarios: (i) a link is added between two core nodes; (ii) a link is added between a core and a border node; and (iii) a link is added between two border nodes.   The nodes belonging to the core and periphery have been identified as having the 10$\%$ largest and smallest degrees, respectively.

The main findings observed in each of these three cases are discussed respectively as follows:

\begin{itemize}
\item[(i)] \emph{core-core}:  Inclusion of a link between nodes 22 and 63.   Though the properties of these nodes
       were substantially changed by the added link, other nodes closer to the periphery of the
       network changed even more intensely.   The nodes in the core tended to change relatively little;
\item[(ii)] \emph{border-border}:  Inclusion of a link between nodes 117 and 194.  The largest topological changes
       were observed for these two nodes, followed by other nodes close to the border of the graph.
       The nodes in the core tended to change relatively little;
\item[(iii)] \emph{core-border}:  Inclusion of a link between nodes 22 e 117.  The largest topological changes
       were again observed respectively to these two nodes, followed by nodes near the border of the graph that also happened to be close to nodes 22 and 117.
\end{itemize}

Overall,  the core nodes, which tend to have larger degree, changed relatively less with
the inclusion of the link.   The border nodes tended to change relatively more. When the change (addition of link) involves nodes with larger degrees, the modification index of the
nodes attached to the altered nodes tend to be less intense, because the cross-relation takes into
account the average of the coincidences, which changes less when there are several edges (higher
degree).

Interestingly, there is not definite correlation between the modification indices between nodes
that are adjacent one another.  Though there are cases, such as node 194 with nodes 46 and 131 in 
Fig (b), in which positive correlation can be observed, there are also several cases in which a node
with a relatively high modification index have neighbors with small modification index values

This tendency turns out to be an interesting property of the cross-relation approach, avoiding relatively
high levels of redundancy that would be otherwise obtained in case the modification indices of
neighboring or nearby nodes were positively (or negatively) correlated.  The lack of this type
of correlation is an ultimate consequence of the fact that the cross-relation takes into account
not only the more immediate neighborhood of each node, but the relationships between the nodes
in all possible neighborhoods.

In order to complement the understanding the impact of modifications of a network on the
\emph{modification index} $\mu$, additional experiments have been performed.  More specifically, not one, but several new links are added to a given network while taking into account the degree of the nodes to which one of the link extremity is attached, while the other extremity is connected to another node chosen with uniform probability.  This experiment has been performed respectively to ER and BA networks, considering 10$\%$ of the nodes with the smallest or largest degrees.  

\begin{figure}[ht]
     \centering
     \begin{subfigure}[b]{0.4\textwidth}
         \centering
 \includegraphics[width=\textwidth]{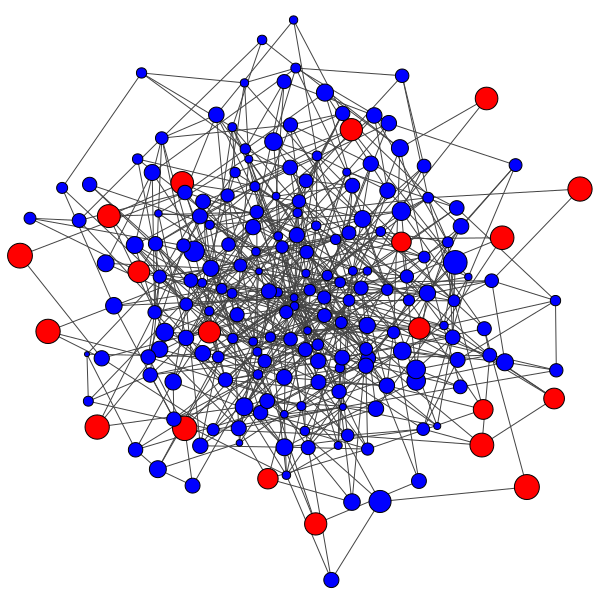}  
         \caption{}
     \end{subfigure} \qquad
     \begin{subfigure}[b]{0.5\textwidth}
         \centering
 \includegraphics[width=\textwidth]{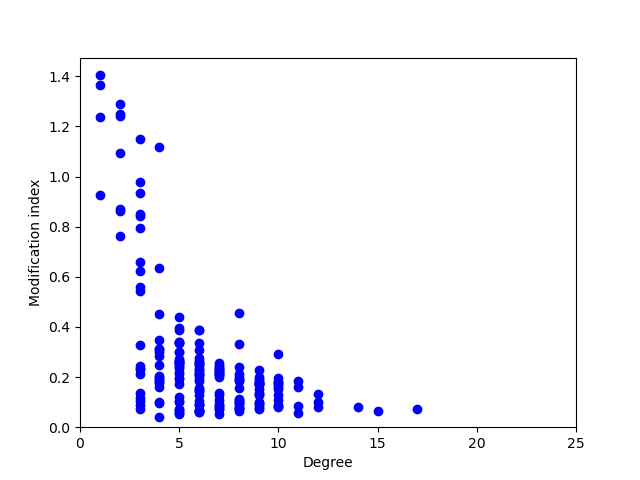}    \\  \vspace{.5cm}
         \caption{}
     \end{subfigure} \\
     \begin{subfigure}[b]{0.4\textwidth}
         \centering
 \includegraphics[width=\textwidth]{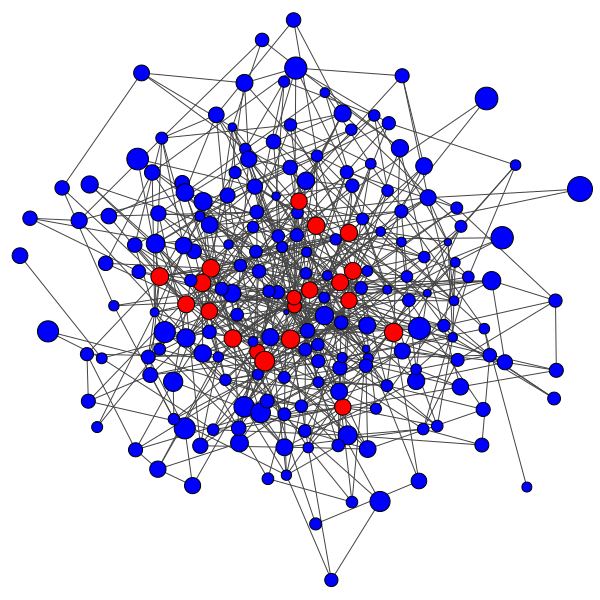}  
         \caption{}
     \end{subfigure}
     \begin{subfigure}[b]{0.5\textwidth}
         \centering
 \includegraphics[width=\textwidth]{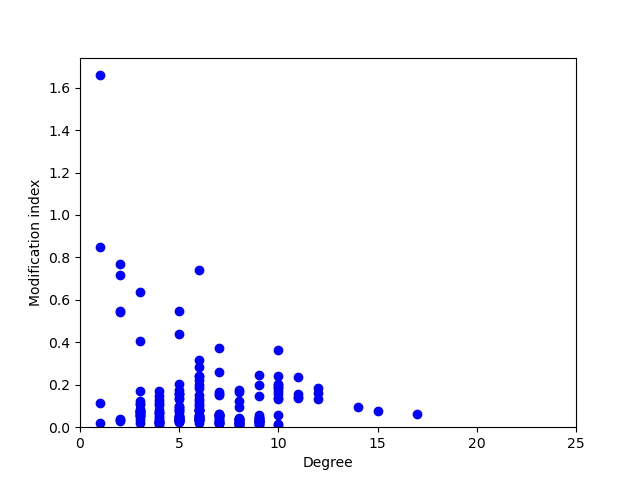}
         \caption{}
     \end{subfigure}
   \caption{Impact of the addition of new edges to the modification index on a ER network. In the first row, the nodes with 10\% lowest degrees received an edge. In the second row, the ones with 10\% highest. Node sizes are defined by the modification index value, linearly scaled by Equation~\ref{eq:linear}.}
  \label{fig:er1}
\end{figure}

Figure~\ref{fig:er1} presents the results of the above described experiments considering ER networks with $N=200$ nodes and average degree $\left<k\right> = 6$.   The cases corresponding to smallest and largest degrees are shown in (a-b) and (c-d), respectively.  More specifically, Figure~\ref{fig:er1}(a) and (c) illustrate an example of the same ER network modified as described above, respectively to new connections taking into account nodes with smallest and largest degrees.  
At the same time, Figures~\ref{fig:er1} (b) and (d) present, respectively to the smallest and largest degrees, the scatterplots of the modification index obtained for each network node in terms of the respective node degrees.  In all cases, the nodes that received new connections are identified in red.  

In the case of modifications of ER networks involving nodes with the smallest degrees, we have that the the nodes resulting in the largest modification indices correspond mostly to the nodes of the periphery of the networks, as illustrated in Figure~\ref{fig:er1}(a).   In their majority, these nodes can be verified to correspond to the nodes  which received new connections.  The remainder nodes were affected in varying manners by the added links, with the most central nodes being more likely to be characterized by smaller modification indices.  However, there are some exceptions to this tendency, including small degree nodes at the border that underwent little modification and more central nodes that resulted in relatively large modifications.  

The tendency of nodes with larger degree to be less influenced by the addition of new connections, observed in the above results, is confirmed by the scatterplot in Figure~\ref{fig:er1}(b).  At the same time, it can be readily verified that the nodes with the smallest degrees can result in a wide range of respective modification values, extending from low to the high index values.   This interesting tendency can be understood to be a consequence of the fact that, by tending to have larger degrees, the topological properties around the core nodes tend to be less susceptible to corresponding modifications implied by the inclusion of a respective link.  In other words, adding a link to a core node that already has a relatively higher number of links will tend to have less impact on the respective hierarchical features than adding a link to a periphery node, which tends to have just a few connections.

\begin{figure}[ht]
     \centering
     \begin{subfigure}[b]{0.45\textwidth}
         \centering
 \includegraphics[width=\textwidth]{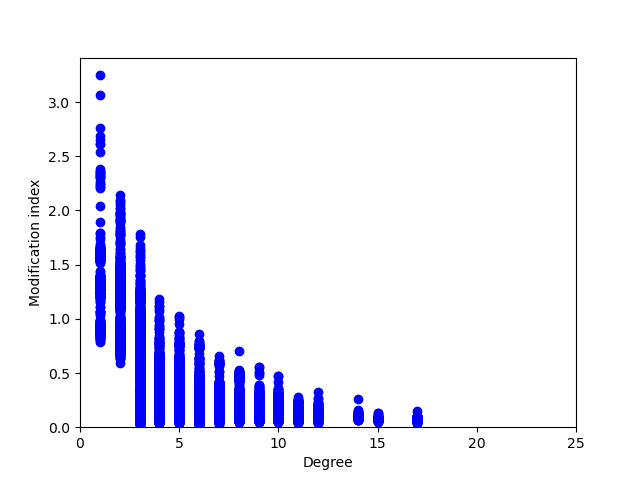}
         \caption{}
     \end{subfigure} \qquad
     \begin{subfigure}[b]{0.45\textwidth}
         \centering
 \includegraphics[width=\textwidth]{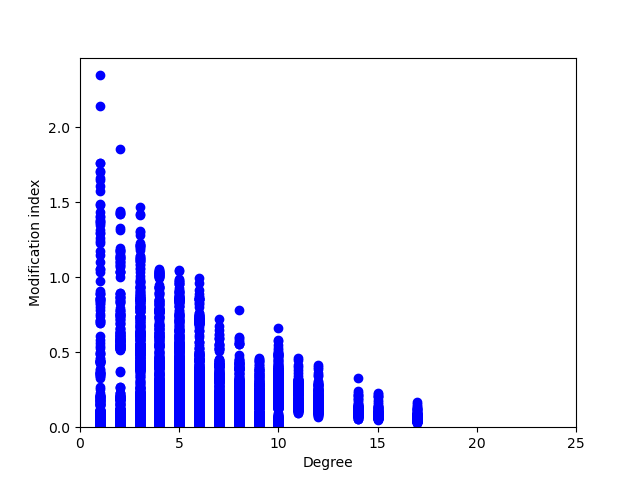}
         \caption{}
     \end{subfigure}
   \caption{Impact of the addition of new edges to the modification index on an ER Network. The figures show the network after addition of a link to the (a) 10\% of the nodes with highest degrees and (b) 10\% of the nodes with lowest degrees. The size of the nodes express the value of the modification index and the red colour identifies the nodes which had a link added. These cases differ from Figures~\ref{fig:er1} (b) and (d) by the number of experiments, which in this case is 100.}
  \label{fig:er2}
\end{figure}
\begin{figure}[ht]
     \centering
     \begin{subfigure}[b]{0.4\textwidth}
         \centering
 \includegraphics[width=\textwidth]{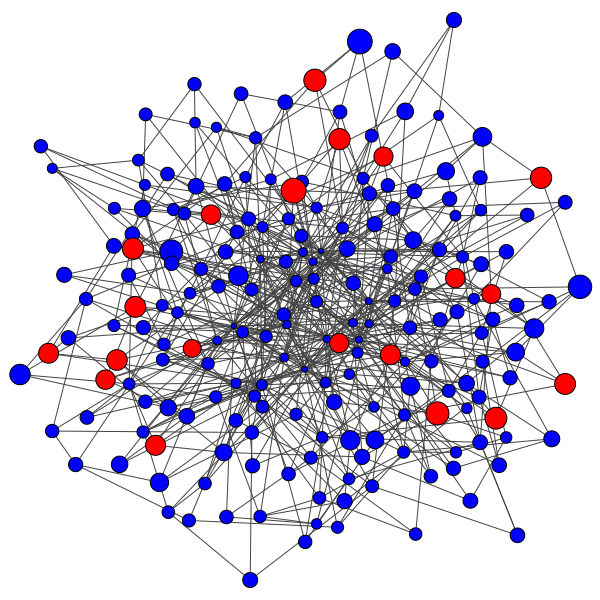}  
         \caption{}
     \end{subfigure} \qquad
     \begin{subfigure}[b]{0.5\textwidth}
         \centering
 \includegraphics[width=\textwidth]{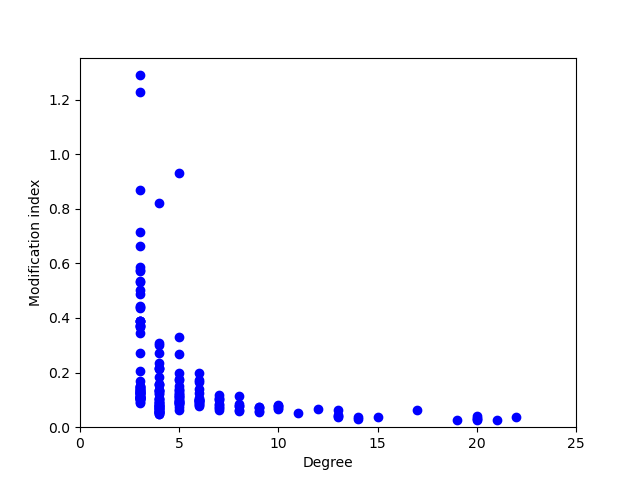}    \\  \vspace{.5cm}
         \caption{}
     \end{subfigure} \\
     \begin{subfigure}[b]{0.4\textwidth}
         \centering
 \includegraphics[width=\textwidth]{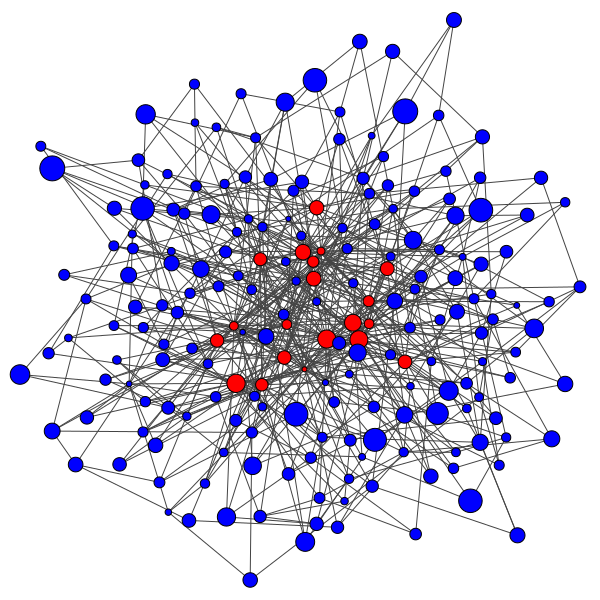}  
         \caption{}
     \end{subfigure}
     \begin{subfigure}[b]{0.5\textwidth}
         \centering
 \includegraphics[width=\textwidth]{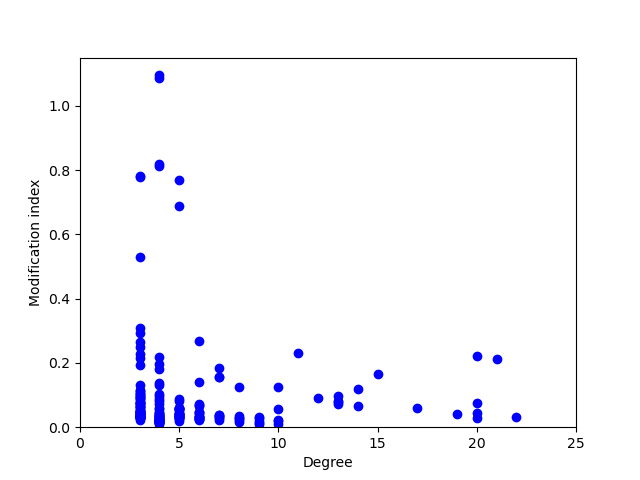}
         \caption{}
     \end{subfigure}
   \caption{Impact of the addition of new edges to the modification index on a BA network. In the first row, the nodes with 10\% lowest degrees received an edge. Opposedly, in the second, the nodes with 10\% highest degrees. Node sizes are defined by the modification index values, linearly scaled by Equation~\ref{eq:linear}.}
  \label{fig:ba1}
\end{figure}

Figure~\ref{fig:ba1}(c) depicts the same ER network as before, but now with new links added to the nodes with the highest degrees.  It can be observed that the nodes with the largest degree, which therefore received new connections, resulted in modification indices with moderate values.
In order to better characterize the relationship between the modification index and the node degree, a total of 100 realizations were performed with the same parameter configuration as the ER example above, i.e.~total of $N=200$ nodes and average degree $\left< k \right>=6$.  The results obtained by adding nodes to the nodes with the largest and smallest degree are presented in Figure~\ref{fig:er2} (a) and (b), respectively.  

In the case of links added to the nodes with the largest degree, shown in Fig.~\ref{fig:er2}(a), 
it can be confirmed from this result that the modification index values decrease steadily with the node degree.  Of particular interest, a void can be observed at the origin of the coordinate system, indicating that the modification indices for node degree 2 and 1 are both lower bound by more than 0.5.  A less abrupt decrease of modification values as the node degree increases can be observed for the case in which links were added to the nodes with the smallest node degrees, shown in  Fig.~\ref{fig:er2}(b).  In addition, no significant gap can be observed along the columns of that scatterplot.

Given that the above experiments concern uniform distribution of links (ER model), it is interesting to repeat those simulations respectively to the BA model-theoretic type of networks.   Figures~\ref{fig:ba1} (a) and (c) show the modification indices (indicated in the node sizes) for the same BA network with $N=200$ and $\left< k \right> = 6$ obtained when links are added to the 10$\%$ nodes with the smallest and largest original degrees.  The scatterplots of modification index in terms of the node degree are shown in (b) and (d), respectively.  

The results in Figures~\ref{fig:ba1}(a) and (c) are mostly similar to the respective counterparts in Figure~\ref{fig:er1} with the main exception that the border nodes in the BA network having undergone relatively large modifications, as indicated by the larger node diameters.
The scatterplots obtained for the BA case, shown in  Figures~\ref{fig:ba1}(b) and (d), are substantially distinct from those obtained for the ER case (Figure~\ref{fig:er1}).   More specifically, though the modification index values also decrease with the node degree, smaller modification values are obtained for smaller degrees.  In addition, gaps can be observed along the columns of the scatterplot.  

\begin{figure}[ht]
  \centering
     \begin{subfigure}[b]{0.45\textwidth}
         \centering
 \includegraphics[width=\textwidth]{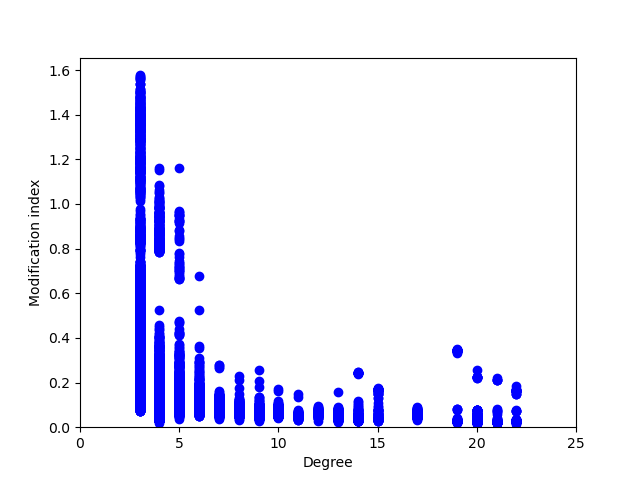}
         \caption{}
     \end{subfigure}
     \begin{subfigure}[b]{0.45\textwidth}
 \includegraphics[width=\textwidth]{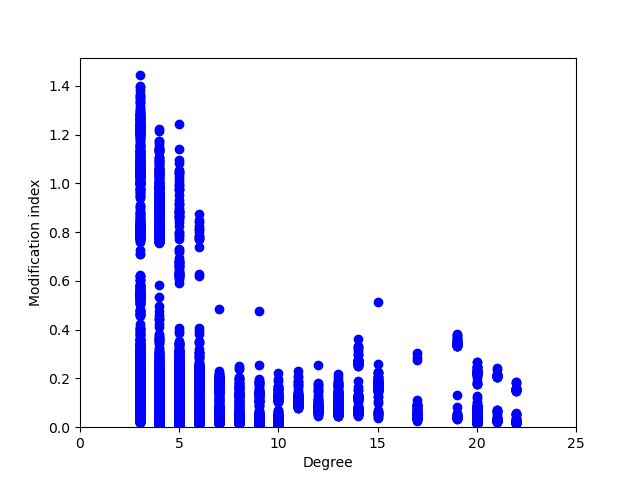}
         \caption{}
     \end{subfigure}
   \caption{Modification index for 100 trials of BA networks. In each experiment, an edge is added to each of the 10\% of lower (higher) nodes are shown on the left (right) figure.}
  \label{fig:ba2}
\end{figure}

Figure~\ref{fig:ba2} presents the scatterplots analogous to those in Figure~\ref{fig:ba1} but considering a total of 100 BA network realizations.  Both obtained scatterplots, respective to links added to the nodes with the largest and smallest degree, are mostly similar, confirming the present of gaps along several
of their columns.

The proposed cross-relation approach has also been compared to an alternative, more direct method in which the features of the nodes at successive neighborhoods are compared in pairwise manner by using the coincidence similarity index.  Please refer to Appendix~\ref{appendix:hierarchical}. More specifically, given the two networks $f$ and $g$ to be compared, and a reference node, a respective signature is obtained consisting of the coincidence similarity values between the features of the two neighborhoods at the same level $\delta$.    The obtained coincidence similarity index, shown in Figure~\ref{fig:hierarchical_er} respectively to the same ER network as in Figure~\ref{fig:er1}, were characterized by the periphery (or core) nodes having markedly distinct values, which contrasts to the results obtained by using the cross-relation approach.  This wide variation of the modification indices obtained by the alternative approach does not adhere to the otherwise expected properties that nodes at the periphery (or core) should be statistically equivalent in ER networks.

\section{A Case-Example: Knowledge Networks}
\
In the previous section we illustrated the application of the modification index to characterize ER and BA networks modified by controlled, specific topological alterations.  In this section, we address the characterization of distinct versions of real-world networks, more specifically two instances (taken in 2014 and 2018) of Physics and Theology knowledge networks obtained from the Wikipedia.  These two versions of each network, taken with a time separation of 4 years, are henceforth called \emph{previous} and \emph{subsequent} instances of the networks.  The procedure adopted to obtain these networks is described in Section~\ref{sec:dataset}.

The number of nodes and edges of each of the obtained networks are presented in Table~\ref{tab:netstats}.

\begin{table}[ht]
\caption{Size of the networks extracted from Wikipedia.}
\centering
\begin{tabular}{|l||l|l|}
\hline
              & $|V|$ & $|E|$ \\ \hline \hline
Physics 2014  & 235 & 578 \\ \hline
Physics 2018  & 273 & 630 \\ \hline
Theology 2014 & 255 & 835 \\ \hline
Theology 2018 & 270 & 927 \\ \hline
\end{tabular}
\label{tab:netstats}
\end{table}

Before proceeding to the analysis of these networks in terms of the modification index, it is interesting to obtain some insight about how the considered networks relate to model-theoretical networks.  In order to do so, we employ the approach described in~\cite{da2022autorrelation}, involving the characterization of the node networks in terms of autorrelation signatures, which are numerically differentiated (in order to enhance their respective differences) and then projected into a PCA space together with model-theoretical networks with the same number of nodes and average degree.  

\begin{figure}[ht]
  \centering
     \begin{subfigure}[b]{0.45\textwidth}
         \centering
 \includegraphics[width=\textwidth]{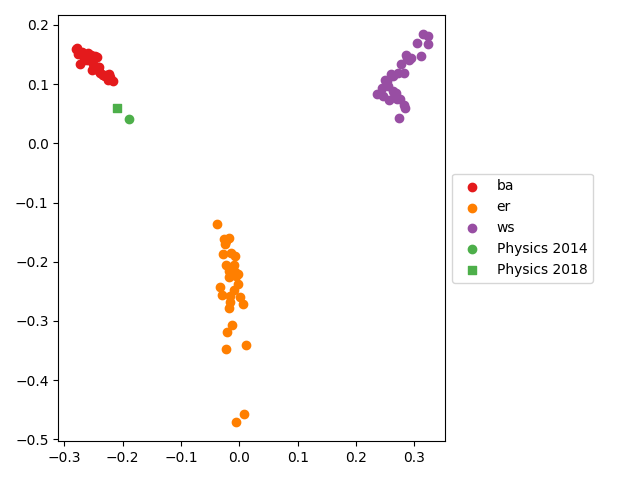}  
 \caption{\hspace*{5em}}
     \end{subfigure}
     \begin{subfigure}[b]{0.45\textwidth}
         \centering
 \includegraphics[width=\textwidth]{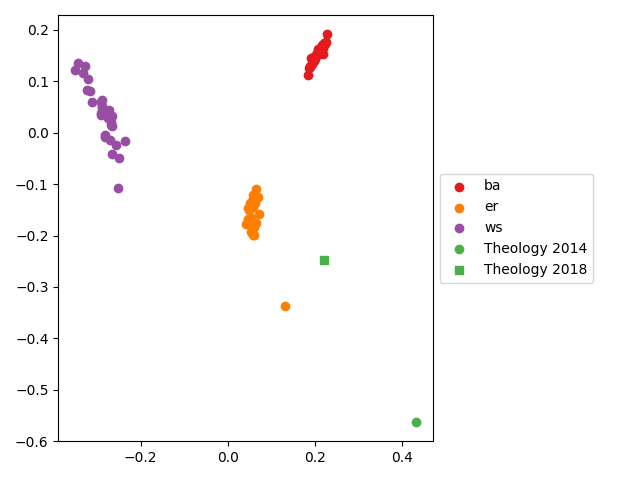}  \\
 \caption{\hspace*{5em}}
     \end{subfigure}
   \caption{The PCA diagrams obtained for the Physics (a) and Theology (b), shown jointly with ensembles of respective model-theoretical networks of types ER, BA, and WS with the same size and average degree.  The previous and subsequent versions of the Physics and Theology networks are shown as green disk and square, respectively.   In the case of the Physics networks, shown in (a), we have that both the previous and subsequent networks resulted markedly close to the BA ensemble of networks.   Contrariwise, as can be observed from the PCA in (b), while the Theology 2014 network resulted close to the ER model, the Theology 2018 network resulted further away from that theoretical network model.  }
  \label{fig:pca}
\end{figure}

Figure~\ref{fig:pca} illustrates the results obtained respectively to the comparison, in respective PCA spaces, of the considered knowledge networks and respective model-theoretical ensembles of ER, BA and WS networks with the same size and average degree.  
In the case of the Physics network, depicted in Figure~\ref{fig:pca}(a), we have that the previous and subsequent instances of the respective knowledge networks resulted substantially close to the ensemble of BA networks, therefore indicating that the considered Physics networks have topology, as far as described by the degree autorrelations, markedly similar to the BA model~\cite{da2022autorrelation}.
Regarding the result obtained for the Theology knowledge networks, shown in Figure~\ref{fig:pca}(b), they mostly constitute \emph{plateaus} that are similar to those obtained for the ER model~\cite{da2022autorrelation}.

\begin{figure}[ht]
  \centering
     \begin{subfigure}[b]{0.45\textwidth}
         \centering
 \includegraphics[width=\textwidth]{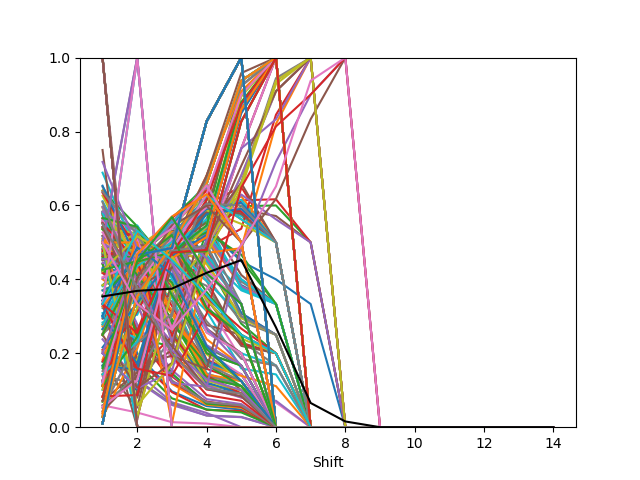}
         \caption{}
     \end{subfigure}
     \begin{subfigure}[b]{0.45\textwidth}
 \includegraphics[width=\textwidth]{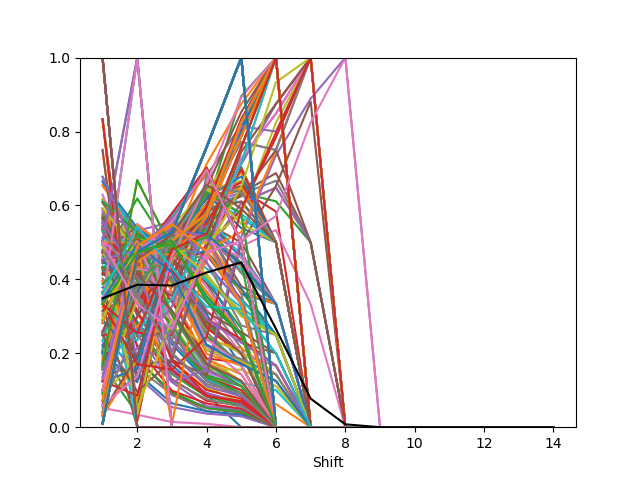}
         \caption{}
     \end{subfigure}\\
     \begin{subfigure}[b]{0.45\textwidth}
         \centering
 \includegraphics[width=\textwidth]{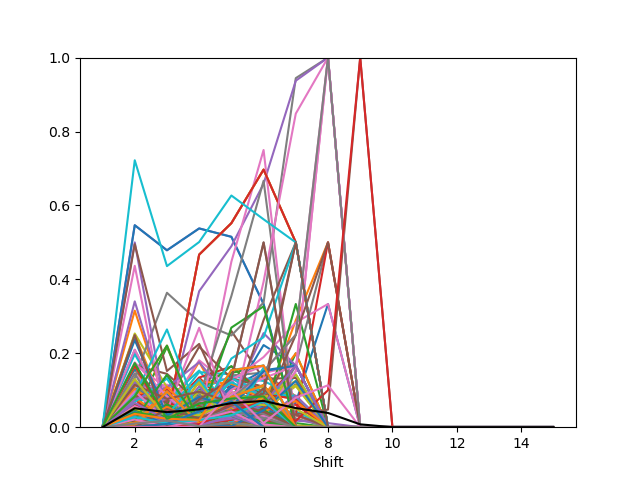}
         \caption{}
     \end{subfigure}
     \begin{subfigure}[b]{0.45\textwidth}
 \includegraphics[width=\textwidth]{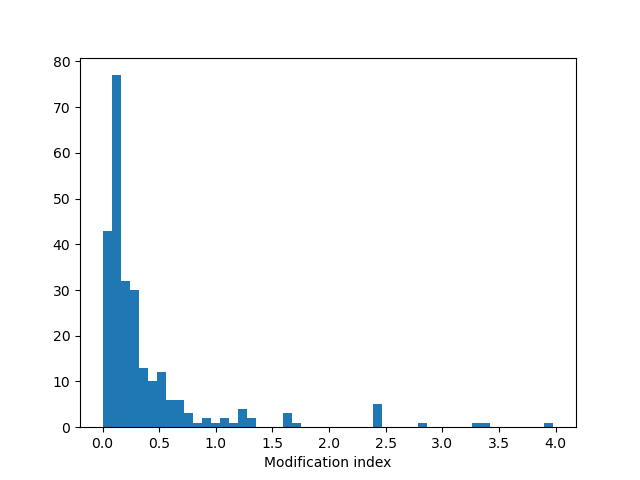}
         \caption{}
     \end{subfigure}
   \caption{The autorrelation signatures of the 2014 (a) and 2018 (b) Physics networks.  The absolute value of the differences between the individual signatures in (a) and (b) are presented in (c).  The modification index of each node is then obtained as corresponding to the area of the respective curves in (c).  The histogram of the modification indices is presented in (d).}
  \label{fig:phys1}
\end{figure}

Figure~\ref{fig:phys1}(a) and (b) present the autorrelation signatures obtained for the 2014 and 2018 instances of the Physics network, respectively.   These two sets of signatures can be verified to be distinct, leading to different average respective signatures (shown in black). Figure~\ref{fig:phys1}(c) illustrates the absolute value of the difference of the signatures in (a) and (b).  Even though the average difference signature is small, comprised in the interval $[0,0.1]$, significant differences between several of the individual signatures can be observed.
The modification index of each node corresponds to the area of the absolute value of the difference of respective signatures taken respectively to the 2014 and 2018 instances of the Theology network.  Figure~\ref{fig:phys1}(d) depicts the histogram of the modification indices obtained for the Physics networks.  It can be observed that most of the modification indices are relatively small, being contained in the interval $[0,1]$.  The most intensely modified nodes can thus be understood to be associated to the histogram samples having modification index larger than 1.  

\begin{figure}[ht]
  \centering
     \begin{subfigure}[b]{0.4\textwidth}
 \includegraphics[width=\textwidth]{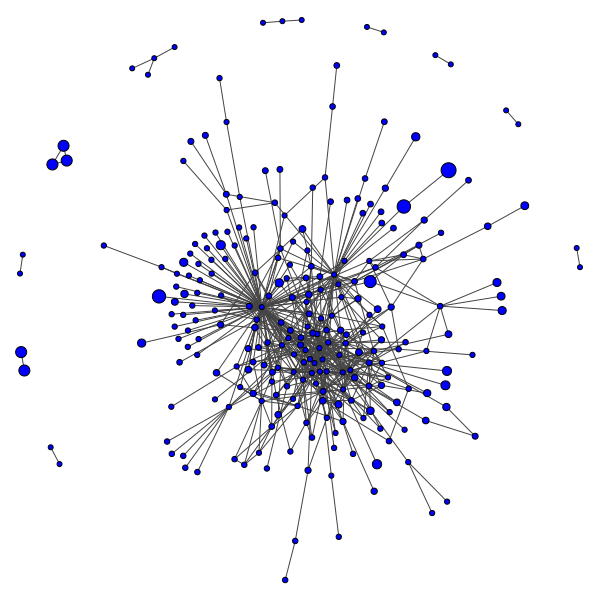}
         \caption{}
 \end{subfigure}
     \begin{subfigure}[b]{0.48\textwidth}
 \includegraphics[width=\textwidth]{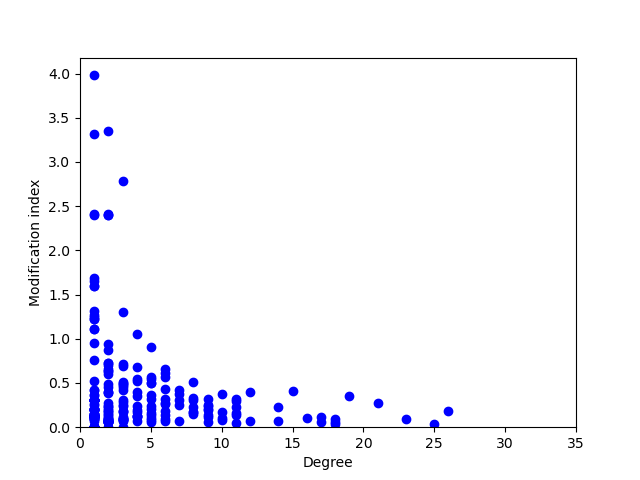}\\
         \caption{}
 \end{subfigure}
     \begin{subfigure}[b]{0.4\textwidth}
 \includegraphics[width=\textwidth]{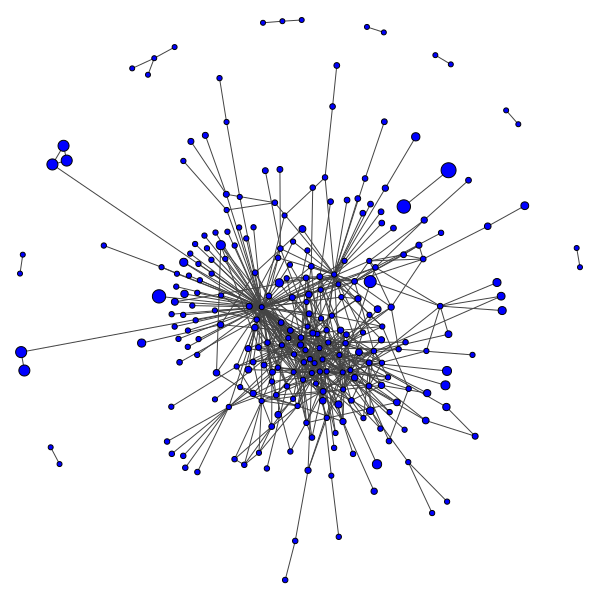}
         \caption{}
 \end{subfigure}
     \begin{subfigure}[b]{0.48\textwidth}
 \includegraphics[width=\textwidth]{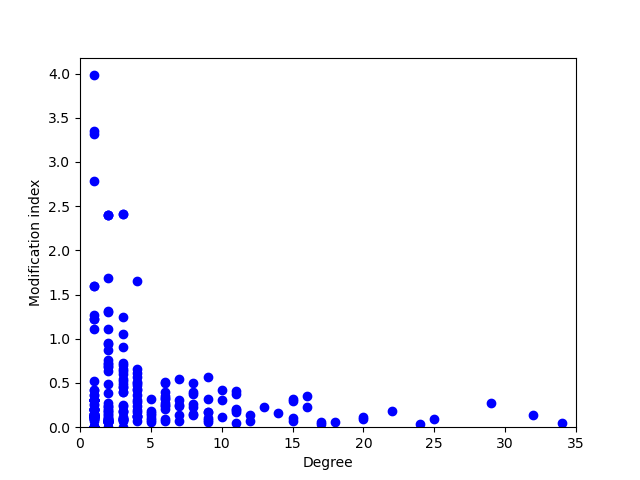}
         \caption{}
 \end{subfigure}
   \caption{The graphs of the 2014 (a) and 2018 (c) instances of the Physics network, with the modification index values of the nodes indicated by the node sizes, logarithmically scaled by Equation~\ref{eq:log}.}
  \label{fig:phys2}
\end{figure}

Several interesting results can be identified from Figure~\ref{fig:phys2}. First, we have that the nodes that underwent the largest topological changes (considering all neighborhoods) are to be found along the periphery of the obtained networks.  Virtually every node at the center of both networks, which tend to have larger degree, were relatively little influenced by the changes in the Physics entries in the Wikipedia from 2014 to 2018.  

\begin{figure}[ht]
  \centering
     \begin{subfigure}[b]{0.45\textwidth}
         \centering
 \includegraphics[width=\textwidth]{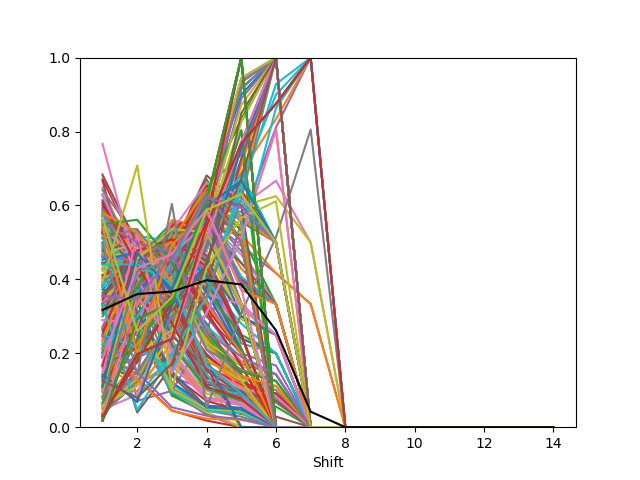}
         \caption{}
     \end{subfigure}
     \begin{subfigure}[b]{0.45\textwidth}
 \includegraphics[width=\textwidth]{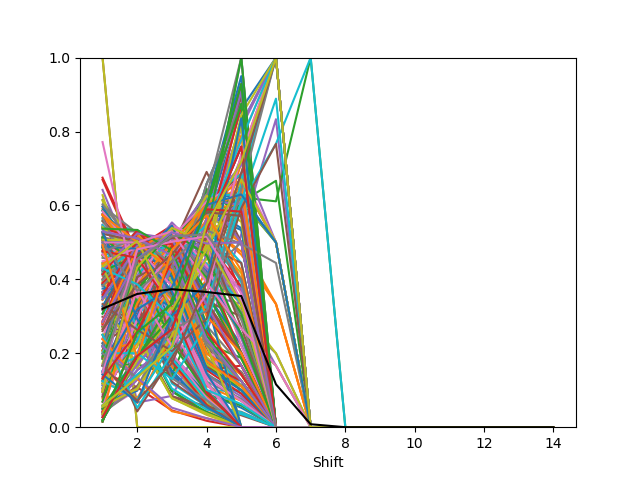}
         \caption{}
     \end{subfigure}\\
     \begin{subfigure}[b]{0.45\textwidth}
         \centering
 \includegraphics[width=\textwidth]{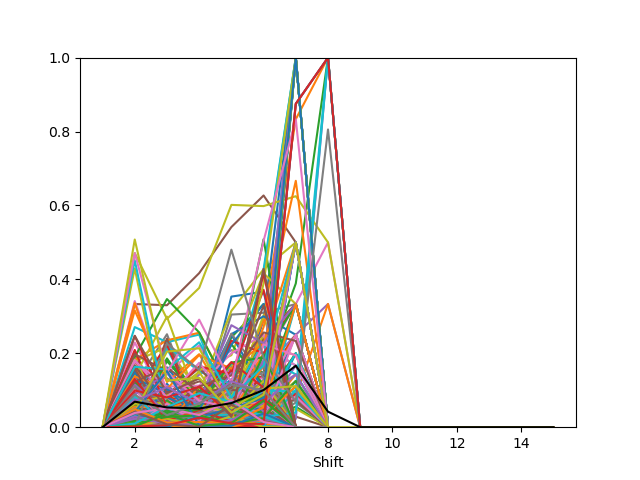}
         \caption{}
     \end{subfigure}
     \begin{subfigure}[b]{0.45\textwidth}
 \includegraphics[width=\textwidth]{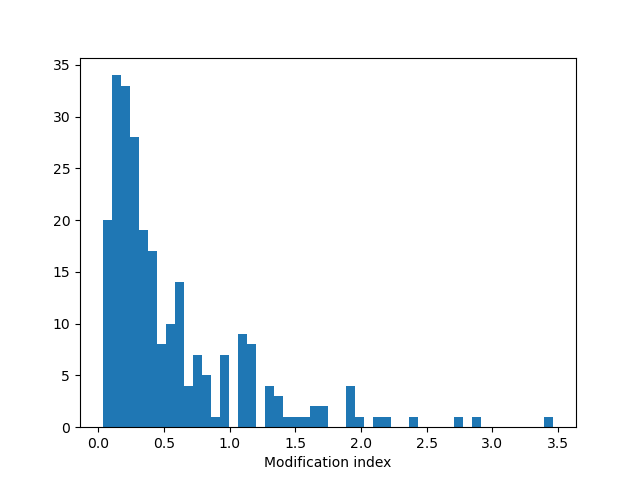}
         \caption{}
     \end{subfigure}
   \caption{The autorrelation signatures of the 2014 (a) and 2018 (b) Theology networks.  The absolute value of the differences between the individual signatures in (a) and (b) are presented in (c).  The modification index of each node is then obtained as corresponding to the area of the respective curves in (c).  The histogram of the modification indices is presented in (d).}
  \label{fig:theo1}
\end{figure}

In order to better characterize the possible relationship between the modification indices and respective node degrees, Figures~\ref{fig:theo1}(b) and (d) present the respective scatterplots.  These scatterplots are similar to those obtained respectively to the theoretical models addressed in Section~\ref{sec:theo} in the sense that smaller modification indices are obtained for nodes with larger degrees, while nodes with smaller degrees can have a wide range of modification index values, varying from 0 to 4.0 in the case of the two instances of the Physics networks.

\begin{figure}[ht]
  \centering
     \begin{subfigure}[b]{0.4\textwidth}
 \includegraphics[width=\textwidth]{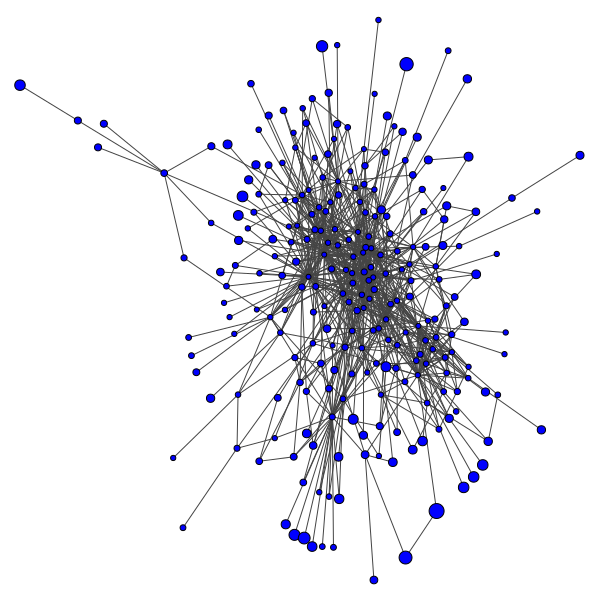}
         \caption{}
 \end{subfigure}
     \begin{subfigure}[b]{0.48\textwidth}
 \includegraphics[width=\textwidth]{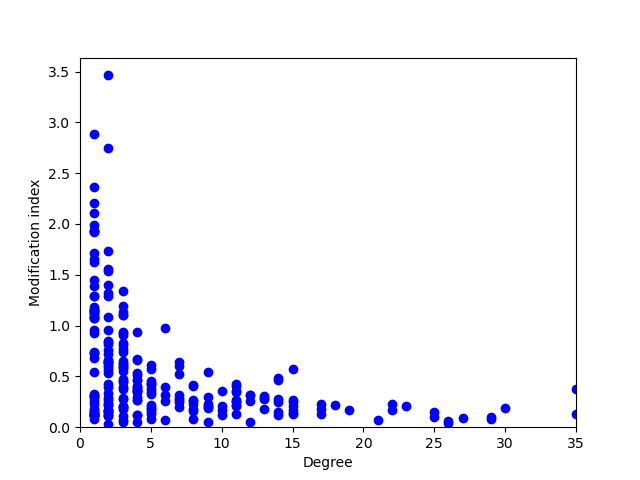}\\
         \caption{}
 \end{subfigure}
     \begin{subfigure}[b]{0.4\textwidth}
 \includegraphics[width=\textwidth]{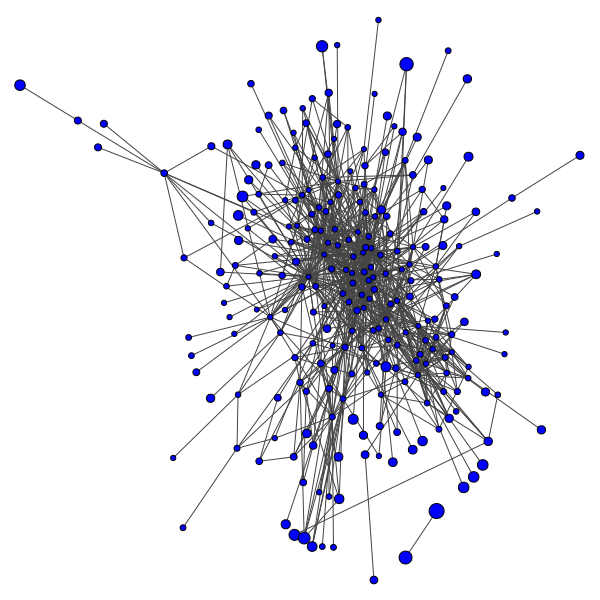}
         \caption{}
 \end{subfigure}
     \begin{subfigure}[b]{0.48\textwidth}
 \includegraphics[width=\textwidth]{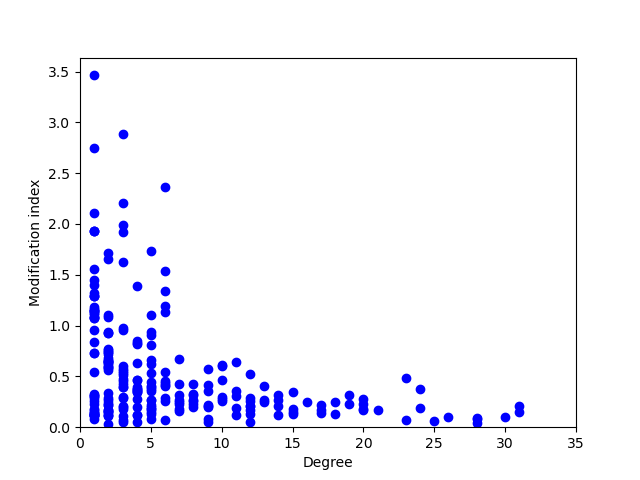}
         \caption{}
 \end{subfigure}
   \caption{The graphs of the 2014 (a) and 2018 (c) instances of the Theology network, with the modification index values of the nodes indicated by the node sizes, logarithmically scaled by Equation~\ref{eq:log}.}
  \label{fig:theo2}
\end{figure}

Figures~\ref{fig:theo2}(a) and (c) show the Theology networks for 2014 and 2018 with the diameter of the nodes indicating, by using the mapping in Equation~\ref{eq:log} the respective modification index values.  

Figure~\ref{fig:theo2} is analogous to Figure~\ref{fig:phys2}, but being now respective to the Wikipedia Theology areas.  As before, noticeable differences can be observed between the respectively obtained signatures, but the two average signatures are now more dissimilar one another.  
At the same time, a significantly smaller average absolute difference signature can be observed in Figure~\ref{fig:phys2}(c), with values comprised in the interval $[0,1]$.  Also in agreement with the results obtained for the Theology area, the nodes characterized by the largest modification indices are again to be found along the periphery of the networks.

\section{Concluding Remarks}

As a great deal of human knowledge finds its way into digital form and the Internet, advances in methodologies for characterizing, studying and modeling this type of data are continuously being required.  As a consequence of their ability to represent virtually every discrete structure, complex networks have been extensively used to represent this type of documents, giving rise to networks in which portions of knowledge are represented as nodes, while references among these portions are expressed as respective edges.   This type of obtained structures has often been understood as belonging to the broader category of \emph{knowledge networks}.

Knowledge networks derived from resources such as the Wikipedia have the interesting property in which their interconnections and nodes can change along time, as new data is continuously incorporated and revised.  The ways in which these changes take place are of special interest, as they can provide clues about how knowledge is dynamically generated, revised and complemented.  In this context, the possibility of prediction of where new nodes and links tend to be included, or removed, constitutes an issue of particular interest.

The present work addressed the possibility to characterize the changes of the network topology around each node, considering two respective snapshots of specific networks between two instants of time.  More specifically, we resourced to the recently described concept of \emph{cross-relation} between two networks, which is based on comparing the topological features of each node with those of the respective nodes at subsequent neighborhoods.  Because of its enhanced selectivity and sensitivity, the similarity coincidence index has been adopted for implementing these comparisons.
More specifically, we combined the cross-relation signatures obtained for each node from both compared networks into a single measurement of respective modification (which has been called the modification index) of the topology of the several neighborhoods around that node.

The application of the proposed approach considered two main cases.  First, in order to infer and validate the proposed methodology, it was applied to the characterization of changes respectively to two theoretical complex network models, namely ER and BA.  The observed results indicated that, in both these models, the core nodes undergo smaller modification of the surrounding topology, with a contrary tendency being obtained for the periphery nodes.  In addition, as could be expected, the nodes belonging presented mostly uniform changes within each of the two groups, though larger heterogeneity has been observed in the case of the BA network.

The second type of application aimed at characterizing topological modifications undergone by knowledge networks concerning the areas of Physics and Theology, obtained from the Wikipedia.
More specifically, we considered two instances, separated by 4 years, of each of these two networks.
The cross-relation signatures obtained for the Physics and Theology networks were found to resemble more closely the signatures that are typically observed for the BA and ER theoretical models.
In addition, in both the Physics and Theology areas, the largest modification indices have been obtained mainly among the respective periphery nodes.  Therefore, this result can be taken as a potentially useful subsidy for predicting alterations in knowledge networks.

The described methodology and results pave the way to several related further developments.  For instance, it would be interesting to consider other areas of knowledge, as well as other types of knowledge networks (e.g.~citations, co-authorship, etc.), or even other types of networks (e.g.~biological, transportation, etc.).  Another possibility would be to consider other time intervals, or even sequences of snapshots along time.

\section{Acknowledgments}
E. K. Tokuda thanks FAPESP (2019/01077-3 and  2021/14310-8) for financial support. 
L. da F. Costa thanks CNPq (307085/2018-0 ) and FAPESP (2015/22308-2) for support.
The work of RL was supported by EPSRC grants EP/V013068/1 and EP/V03474X/1. 

\bibliographystyle{unsrt}
\bibliography{main}

\appendix
\section{Direct Comparison of Hierarchical Features}
\label{appendix:hierarchical}

Given two networks $f$ and $g$ with aligned nodes, an alternative to obtain signatures characterizing the similarity between the topological structure along hierarchies defined by respective reference nodes consists in comparing the features, in pairwise fashion, between the nodes in the same hierarchical levels along the two networks.  

\begin{figure}
    \centering
    \includegraphics[width=.9\textwidth]{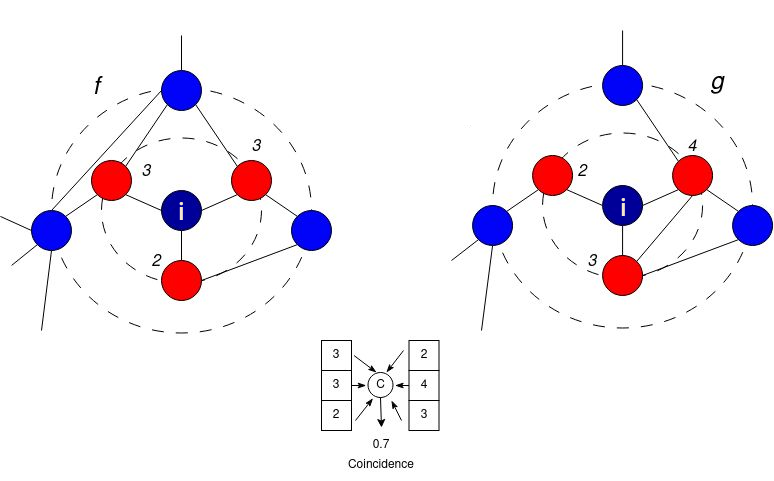}
    \caption{Comparison of two networks $f$ and $g$ considering the first hierarchical degree with the center node $i$ as reference.}
    \label{fig:hierarchical_diagram}
\end{figure}

Figure~\ref{fig:hierarchical_diagram} illustrates this method respectively to a specific pair of networks $f$ and $g$ and the first neighborhoods ($\delta=1$) respective to the reference node $i$. The node features are taken as corresponding to the respective degrees.  Also illustrated is the comparison, by using the coincidence similarity index, of the node degrees of the nodes in the first neighborhood of networks $f$ and $g$, which leads to the value $0.7$.

This method has its performance quantified in terms of the same experiments as described in Section~\ref{sec:theo}.  Starting from a reference network, a set of interest ($m$) nodes is defined. From each of these nodes, a new neighbour is added in uniformly random manner. The obtained network with $m$ additional edges is then compared with the respective original network by  using the above described approach. 

\begin{figure}[ht]
  \centering
     \begin{subfigure}[b]{0.45\textwidth}
         \centering
 \includegraphics[width=\textwidth]{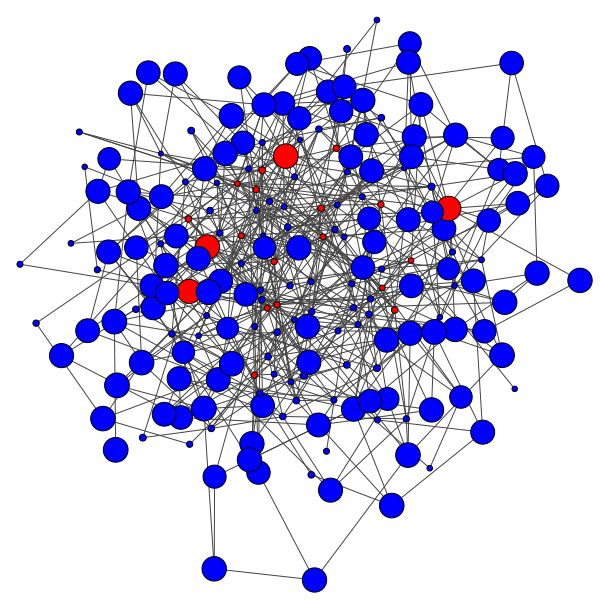}  
 \caption{\hspace*{5em}}
     \end{subfigure}
     \begin{subfigure}[b]{0.45\textwidth}
         \centering
 \includegraphics[width=\textwidth]{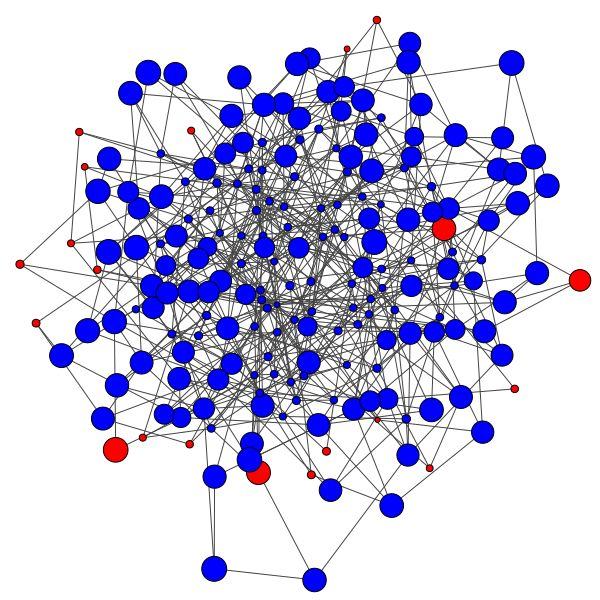}  \\
 \caption{\hspace*{5em}}
     \end{subfigure}
   \caption{Method applied to an ER network corresponding to the addition of a single edge to the (a) core nodes and (b) periphery nodes.}
  \label{fig:hierarchical_er}
\end{figure}

In Figure~\ref{fig:hierarchical_er}, we show an ER network and two sets of reference nodes (shown in red) corresponding respectively to the core (Figure~\ref{fig:hierarchical_er}(a)) and the periphery (Figure~\ref{fig:hierarchical_er}(b)) of the network. The size of the nodes corresponds to the coincidence index between the node degrees along the successive neighborhoods.  The obtained core and periphery nodes can be observed to be characterized by markedly distinct sizes within their respective groups, therefore indicating great dispersion of coincidence index values, in contrast to the otherwise expected uniformity of topological properties among these two types of nodes.  This result therefore indicates that the direct comparison of neighborhoods does not provide a stable approach to characterizing the topological alterations undergone by each modified node along the respective hierarchies, as had been obtained for the cross-relation approach as described in Section~\ref{sec:theo}.

\end{document}